\documentclass[prd,twocolumn,showpacs,superscriptaddress,nofootinbib,floatfix,showkeys,10pt]{revtex4-1}
\usepackage{graphicx}
\usepackage{amsmath}
\usepackage{bm}
\usepackage{yhmath}
\usepackage{mathtools}
\usepackage{wasysym}
\usepackage[colorlinks,citecolor=violet,urlcolor=violet,linkcolor=blue]{hyperref}
\usepackage{color}
\usepackage{cases}
\usepackage{subfigure}
\usepackage{times}
\usepackage{dcolumn,booktabs,bm}
\usepackage{slashed}
\usepackage{amsfonts,amssymb,stmaryrd,latexsym,amsmath}
\usepackage{textcomp}
\usepackage{multirow}
\usepackage{cancel}
\usepackage{array}
\usepackage{orcidlink}
\usepackage{physics}

\def\SU3{{\text{SU(3)}_{\rm F}}}

\def \pcs4338{{P_{\psi s}^\Lambda(4338)^0}}

\renewcommand{\arraystretch}{1.8}
\allowdisplaybreaks

\begin{document}
	
	\title{\textcolor{violet}{The charge and magnetic radii of the nucleons from the generalized parton 
	                          distributions}}
	
	\author{The MMGPDs\footnote{Modern Multipurpose GPDs} Collaboration:\\
	        Muhammad Goharipour\,\orcidlink{0000-0002-3001-4011}}\email{muhammad.goharipour@ipm.ir}    
    \thanks{Corresponding author}
	\affiliation{Department of Physics, University of Tehran, North Karegar Avenue, Tehran 14395-547, Iran}
	\affiliation{School of Particles and Accelerators, Institute for Research in Fundamental 
		         Sciences (IPM) P. O. Box 19395-5531, Tehran, Iran}
	
	\author{Fatemeh Irani\,\orcidlink{0009-0008-3447-6361}}\email{f.irani@ut.ac.ir}
	\affiliation{Department of Physics, University of Tehran, North Karegar Avenue, Tehran 14395-547, Iran}
	
	\author{Hadi Hashamipour\,\orcidlink{0009-0003-6689-9637}}\email{hadi.hashamipour@lnf.infn.it}
	\affiliation{Istituto Nazionale di Fisica Nucleare, Gruppo collegato di Cosenza, I-87036 Arcavacata di Rende, Cosenza, Italy}
	
	\author{K. Azizi\,\orcidlink{0000-0003-3741-2167}}\email{kazem.azizi@ut.ac.ir} 
	\affiliation{Department of Physics, University of Tehran, North Karegar Avenue, Tehran 14395-547, Iran}
	\affiliation{School of Particles and Accelerators, Institute for Research in Fundamental 
		         Sciences (IPM) P. O. Box 19395-5531, Tehran, Iran}
	\affiliation{Department of Physics, Dogus University, Dudullu-\"Umraniye, 34775 Istanbul, 
	             T\"urkiye}

	
	\begin{abstract}
	
The proton-radius puzzle refers to the discrepancy observed in measurements of the proton's charge radius when using different methods. This inconsistency has prompted extensive research and debate within the physics community, as it challenges the understanding of quantum electrodynamics and the fundamental properties of protons. In the present study, we determine the charge and magnetic radii of the proton and neutron through a global analysis of the generalized parton distributions (GPDs) at zero skewness. We emphasize the importance of a simultaneous analysis of all available experimental data related to nucleon radii, rather than relying on individual experiments, specific observables, or limited kinematic regions. This comprehensive approach ensures robust and consistent results, avoiding values that are either too small or too large. Our analysis yields the following results: $ r_{pE} = 0.8558 \pm 0.0135~\textrm{fm} $, $ r_{pM} = 0.8268 \pm 0.0533~\textrm{fm} $, $ \left<r_{nE}^2\right> = -0.1181 \pm 0.0270~\textrm{fm}^2 $, and $ r_{nM} = 0.8367 \pm 0.0845~\textrm{fm} $.
		
	\end{abstract}
	
	
	\maketitle
	
	\thispagestyle{empty}
	
	\textit{\textbf{\textcolor{violet}{Introduction}}}~~
Understanding the internal structure of nucleons, encompassing protons and neutrons, remains a fundamental pursuit in modern nuclear and particle physics. Central to this investigation are the charge and magnetic radii of the nucleons, critical parameters that reveal the spatial distribution of their electric charge and magnetic moment~\cite{Karr:2020wgh}. These properties not only illuminate the dynamics within atomic nuclei but also serve as crucial benchmarks for testing theoretical models, particularly quantum electrodynamics (QED) and quantum chromodynamics (QCD).

The charge radius of a nucleon defines the spatial extent over which its electric charge is distributed, while the magnetic radius characterizes the distribution of its magnetic moment. Precise measurements of these radii serve as stringent tests for theories such as chiral effective field theory~\cite{Gil-Dominguez:2023hku} and lattice QCD calculations~\cite{Park:2021ypf}. 
Electromagnetic scattering experiments~\cite{Gao:2021sml,Xiong:2023zih}, utilizing high-energy electron beams, have traditionally been employed to determine these radii by measuring the cross section of the elastic scattering of electrons off nucleons in terms of the negative of the square
of the four-momentum transfer $ Q^2=-t $. In fact, it is now well established that the scattering experiments access to the charge and magnetic radii of the nucleons through the slope of the electromagnetic form factors (FFs) at $ Q^2=0 $.
In parallel, spectroscopic techniques, such as hydrogen (H) or muonic hydrogen ($ \mu $H) spectroscopy, provide an alternative avenue for probing nucleon structure~\cite{Pohl:2010zza,Grinin:2020txk}. These methods involve studying the energy levels and transitions within exotic atoms, while scattering experiments offer direct access to the nucleon's radii via electromagnetic interactions, spectroscopy provides a unique perspective through atomic energy level measurements.
	
In 2010, the CREMA collaboration performed a challenging muonic hydrogen spectroscopy experiment, an improvement of the hydrogen spectroscopy method wherein the electron is replaced by a muon, and
reported a value of the proton charge radius ($ r_p $) which was in disagreement with previous determinations from both ordinary hydrogen spectroscopy and elastic scattering experiments~\cite{Pohl:2010zza}. This finding, which was referred to as the ``proton-radius puzzle", has been a critical subject to investigate in both experimental and theoretical particle physics for more than a decade~\cite{Pohl:2013yb,Bernauer:2014cwa,Carlson:2015jba,Hill:2017wzi,Peset:2021iul,Lin:2023fhr}. However, two recent measurements from both hydrogen spectroscopy~\cite{Bezginov:2019mdi} and electron-proton ({\it ep}) elastic scattering from PRad experiment~\cite{Xiong:2019umf} at Jefferson Laboratory (JLab) are in agreement with the muonic hydrogen spectroscopy results which lead to smaller values for $ r_p $ of about 0.83 fm. Note that the PRad result is the first measurement of $ r_p $ obtained from elastic scattering experiments that is in agreement with the muonic hydrogen spectroscopy results. Actually, the results from previous elastic scattering experiments are larger than the PRad result even up to 5 standard deviations. The reason has been attributed to the fact that the PRad experiment covers the measurements of the {\it ep} elastic scattering cross section (and hence the electric FF) at very small values of $ Q^2 $. Nevertheless, a recent study~\cite{Ridwan:2023ome} shows that analyzing PRad data besides other data belonging to larger values of $ Q^2 $ leads to a larger value of $ r_p $ which is again in disagreement with the muonic hydrogen spectroscopy results.

The precise measurements of the nucleon's charge and magnetic radii are also important for the global analysis of the generalized parton distributions (GPDs) at zero longitudinal momentum transfer $ \xi $, known as skewness, since they can provide crucial constraints on GPDs at very small $ Q^2 $~\cite{Hashamipour:2021kes,Hashamipour:2022noy,Goharipour:2024atx}. Representing correlations between the longitudinal momentum and the transverse position and of partons inside the nucleon,
GPDs are nonperturbative entities which can provide a 3D description of nucleons~\cite{Muller:1994ses,Radyushkin:1996nd,Ji:1996nm,Ji:1996ek,Burkardt:2000za,Goeke:2001tz,Diehl:2003ny,Belitsky:2005qn,Boffi:2007yc,Diehl:2015uka}. So, in this context, they have a significant advantage compared with ordinary parton distribution functions (PDFs) that describe only the longitudinal momentum distribution of partons inside hadrons. This is why the GPDs topic has received a lot of attention in recent years~\cite{Guidal:2004nd,Diehl:2013xca,Berthou:2015oaw,Kumericki:2016ehc,Constantinou:2020hdm,Kriesten:2021sqc,Guo:2022upw,Duplancic:2022ffo,Guo:2023ahv,Duplancic:2023kwe,Kaur:2023lun,Liu:2024umn,Riberdy:2023awf,Luan:2024vgv,Bhattacharya:2024qpp,Arami:2024qsu,Goloskokov:2024egn,Bhattacharya:2024wtg,Thakuria:2024nyv,Irani:2023lol}.  Note that this list is not exhaustive, as it is nearly impossible to provide a fully comprehensive compilation of all relevant GPD-related studies within the scope of this paper. At zero skewness, GPDs can be determined from a simultaneous analysis of the experimental data from different kinds of elastic scattering processes including the elastic lepton-nucleon scattering~\cite{Goharipour:2024atx}, elastic (anti)neutrino-nucleon scattering~\cite{Irani:2023lol}, and wide-angle Compton scattering (WACS)~\cite{Hashamipour:2022noy}. An important property of GPDs is that their first moments are related to different types of nucleon FFs and hence the nucleon's radii. Therefore, it is of significant interest to see how GPDs determined from a simultaneous analysis of the electromagnetic FFs (or elastic scattering cross sections) predict the charge and magnetic radii of the nucleons. Are the obtained values for radii in agreement with PRad and muonic hydrogen spectroscopy results? 

It is worth noting that the determination of GPDs at zero skewness through the analysis of electromagnetic FFs and their impact on nucleon radii has been performed before, for example, in Refs.~\cite{Guidal:2004nd,Diehl:2013xca}. However, such analyses are based on fewer experimental data points, limited to elastic electron-nucleon scattering, and employed simpler ansatz. The present study utilizes a wide range of experimental data, including both elastic electron-nucleon scattering and quasielastic antineutrino-nucleon scattering. Our analysis is based on a comprehensive data set from the more recent YAHL18 analysis~\cite{Ye:2017gyb} and incorporates new GMp12 data from JLab~\cite{Christy:2021snt} as well as neutrino data from the PRad experiment~\cite{Xiong:2019umf}.

In the present study, we first perform a standard $ \chi^2 $ analysis of a wide range of the elastic electron-nucleon scattering as well as the antineutrino-nucleon scattering experimental data and determine simultaneously the unpolarized valence GPDs $ H_v^q $ and $ E_v^q $ with their uncertainties. Then, using the extracted GPDs, we calculate the charge and magnetic radii of both proton and neutron and compare them with the corresponding ones obtained from various elastic scattering experiments and hydrogen spectroscopy.

	
	\textbf{\textit{\textcolor{violet}{Formalism}}}~~
As mentioned before, the charge and magnetic radii of the nucleons are directly related to the electric and magnetic FFs, $ G_E^j(Q^2) $ and $ G_M^j(Q^2) $, which encode the internal charge and magnetization distributions. Here $ j=p,n $ where $ p $ and $ n $ denote the proton and neutron, respectively. To be more precise, the mean squared of the charge and magnetic radii of the
nucleons, $ r_{jE} $ and $ r_{jM} $, can be obtained from the slope of the electromagnetic FFs at $ t=0 $ as follows~\cite{Hashamipour:2021kes}
\begin{align}
\left<r_{jE}^2\right>= \left.  6 \dv{G_E^j}{t} \right|_{t=0} \,, ~~~~~~
\left<r_{jM}^2\right>= \left.  \frac{6}{\mu_j} \dv{G_M^j}{t} \right|_{t=0}\,,
\label{Eq1}
\end{align}
where $ \mu_j $ is the magnetic moment of the nucleon. The electromagnetic FFs $ G_E $ and $ G_M $ which are commonly called Sachs FFs are expressed in terms of the Dirac and Pauli FFs of the nucleon, $ F_1 $ and $ F_2 $, 
\begin{align}
\label{Eq2}
G_M(t) &= F_1(t) + F_2(t) \,, \nonumber \\ 
G_E(t) &= F_1(t) + \frac{t}{4m^2} F_2(t) \,,
\end{align}
where $ m $ is the nucleon's mass. 

Experimentally, the Sachs FFs can be extracted from the measurements of the cross section of the elastic electron-nucleon scattering~\cite{Ye:2017gyb} that is often presented as the reduced cross section
\begin{eqnarray}
\label{Eq3}
\nonumber
\sigma_R &=& \epsilon G_E^2(t) + \tau G_M^2(t)\,, \\
 &=& G_M^2(t)[\tau +\epsilon {\rm RS}(t)/\mu_j^2]\,,
\end{eqnarray}
where $\epsilon$ and $\tau$ are the dimensionless kinematic variables and
$ {\rm RS}=(\mu G_E/G_M)^2 $ is the normalized Rosenbluth slope~\cite{Christy:2021snt}.
The method that provides the separate determination of $ G_E $ and $ G_M $ is called the Rosenbluth separation in which the unpolarized elastic electron-nucleon scattering is considered. However, there is another method in which one can use the correlation between the polarizations of the electron and nucleon to extract Sachs FFs. Although such a method has the advantage that it is less sensitive to two-photon exchange (TPE) corrections, one can only access the ratio of Sachs FFs in this method.

Theoretically, the Dirac and Pauli FFs can be obtained from the unpolarized valence GPDs $ H_v^q $ and $ E_v^q $ at zero skewness ($ \xi=0 $). Here $ q $ denotes the up, down, and strange quarks, though we can neglect the strange contribution. Considering the proton case, we have the following sum rules~\cite{Guidal:2004nd,Diehl:2013xca}
\begin{align}
F^p_1(t)=\sum_q e_q F^q_1(t)=\sum_q e_q \int_{0}^1 dx\, H_v^q(x,\mu^2,t)\,, \nonumber \\ 
F^p_2(t)=\sum_q e_q F^q_2(t)=\sum_q e_q \int_{0}^1 dx\, E_v^q(x,\mu^2,t)\,,
\label{Eq4}
\end{align}
where $ x $ represents the longitudinal momentum fraction of the nucleon carried by quark $ q $ with electric charge $ e_q $, and $ \mu^2 $ stands for the factorization scale at which the quarks are resolved.
Note that the related formulas for the neutron FFs $ F_1^n $ and $ F_2^n $ can be obtained by assuming the isospin symmetry, i.e., $ u^p=d^n, d^p=u^n $.

According to the above equations, by measuring the reduced cross section $ \sigma_R $ or equivalently the electromagnetic FFs $ G_E $ and $ G_M $, one can determine the unpolarized valence GPDs $ H_v^q $ and $ E_v^q $ at zero skewness by performing a standard $ \chi^2 $ analysis and considering suitable parameterizations for GPDs~\cite{Hashamipour:2021kes,Hashamipour:2022noy,Goharipour:2024atx}. On the other hand, one can calculate the nuecleon's charge and magnetic radii using Eq.~(\ref{Eq1}) and extracted GPDs.
This approach to calculate the nucleon's radii is really of interest because GPDs are determined from a simultaneous analysis of a wide range of experimental data from the elastic electron scattering off both proton and neutron. While the experimental measurements of the nucleon's radii from elastic scattering experiments are usually based on the fit to certain data (with limited kinematic coverage) and to one kind of FFs solely. For example, in the PRad experiment~\cite{Xiong:2019umf}, the valued of $ r_p=0.831 $ fm has been obtained by analyzing just the electric FF data in the $ t $ range of $ 0.0002 \lesssim -t \lesssim 0.06 $ GeV$^2$. So, the values of the radii of the nucleons determined from the global analysis of GPDs can be crucial in the context of the proton-radius puzzle.

The first step to extract GPDs from the $ \chi^2 $ analysis of the related experimental data is to choose a suitable parametrization for them. It is now well established that the following ansatz which relates GPDs to ordinary PDFs works well~\cite{Hashamipour:2021kes,Hashamipour:2022noy,Goharipour:2024atx,Diehl:2013xca} 
\begin{align}
H_v^q(x,\mu^2,t)= q_v(x,\mu^2)\exp [tf_v^q(x)]\,,  \nonumber \\ 
E_v^q(x,\mu^2,t)= e_v^q(x,\mu^2)\exp [tg_v^q(x)]\,,
\label{Eq5}
\end{align}
where $ q_v(x,\mu^2) $ represents the unpolarized valence PDF for quark flavor $ q $. As in our previous analysis~\cite{Goharipour:2024atx}, utilizing the \texttt{LHAPDF} package~\cite{Buckley:2014ana}, 
we take PDFs from the \texttt{NNPDF} analysis~\cite{NNPDF:2021njg} at the next-to-leading order (NLO) and scale $ \mu=2 $ GeV. Note that the above ansatz satisfies the requirement that GPDs are reduced to PDFs at the so-called forward limit ($ t=0 $ and $ \xi=0 $). Since $ e_v^q(x,\mu^2) $ is not available from the analysis of the high-energy experimental data, we have to determine it from the fit to data too. To this aim, we consider the following parametrization~\cite{Hashamipour:2021kes,Hashamipour:2022noy,Goharipour:2024atx,Diehl:2013xca} 
\begin{equation}
\label{Eq6}
e_v^q(x,\mu^2)=\kappa_q N_q x^{-\alpha_q} (1-x)^{\beta_q} (1+\gamma_q\sqrt{x})\,,
\end{equation}
where $ \kappa_u=1.67 $ and $ \kappa_d=-2.03 $ are the magnetic moments of the up and down quarks, respectively, in the units of nuclear magneton. They have been extracted from the corresponding values of the magnetic moments of proton and neutron taken from~\cite{ParticleDataGroup:2024cfk}.
The profile functions $ f_v^q(x) $ and $ g_v^q(x) $ in Eq.~(\ref{Eq5}) describe the damping of GPDs with respect to PDFs as the value of $ t $ increases. We use the following form to parameterize them as suggested in~\cite{Diehl:2013xca}
\begin{equation}
\label{Eq7}
{\cal F}(x)=\alpha^{\prime}(1-x)^3\log\frac{1}{x}+B(1-x)^3 + Ax(1-x)^2.
\end{equation}

In this way, for each quark flavor, the unknown free parameters that must be determined from the $ \chi^2 $ analysis of the experimental data are  $ \alpha $, $ \beta $, $ \gamma $, $ \alpha^{\prime} $, $ A $, and $ B $. Note that the experimental error associated with each measured data point that is used in the calculation of the $ \chi^2 $ function is obtained by adding the systematic and statistical errors in quadrature. As in Refs.~\cite{Hashamipour:2021kes,Hashamipour:2022noy,Goharipour:2024atx}, the CERN program library \texttt{MINUIT}~\cite{James:1975dr} is used for performing the minimization procedure and determining the optimum values of the unknown parameters. Also, the standard Hessian approach~\cite{Pumplin:2001ct} with $ \Delta \chi^2=1 $ is used to calculate uncertainties. Another point that should be mentioned is that we impose the condition $ g_v^q(x) < f_v^q(x) $ in the main body of the fit program to satisfy the positivity property of GPDs in a wide range of the $ x $ values. Moreover, in order to determine the optimum values of the unknown parameters, we adopt the parametrization scan procedure in which the final parametrization of each distribution is obtained systematically. See Refs.~\cite{Hashamipour:2021kes,Hashamipour:2022noy,Goharipour:2024atx} to get more information about our phenomenological framework. 

Now it is time to discuss the experimental data sets that we use in our analysis of GPDs. The main body of the experimental data that we consider in the present study is as our recent analysis~\cite{Goharipour:2024atx} where we have investigated the impact of JLab data at high values of $ Q^2 $ on the extracted GPDs at zero skewness. These data include the world $ R^p=\mu_p G_E^p/G_M^p $ polarization, $ G_E^n $, and $ G_M^n/\mu_n G_D $ measurements from the YAHL18 analysis~\cite{Ye:2017gyb}, the data of $ G_E^p $ from AMT07~\cite{Arrington:2007ux} and Mainz~\cite{A1:2013fsc} and the measurements of the reduced cross-section $ \sigma_R $ from JLab experiment~\cite{Christy:2021snt}, namely GMp12. Overall, these data cover the  $ -t $ range from 0.007 to 15.76 GeV$ ^{2} $. As mentioned before, in the present study we also add the PRad data~\cite{Xiong:2019umf} which cover the $ -t $ range from 0.00022 to 0.05819 GeV$ ^{2} $ (note that we use both 1.1 GeV and 2.2 GeV data). The total number of data with and without PRad data is 348 and 277, respectively.

	
	\textbf{\textit{\textcolor{violet}{Analysis and Results}}}~~ 
In the following, we first determine the unpolarized valence GPDs $ H_v^q $ and $ E_v^q $ at zero skewness by performing three analyses. In the first analysis, we consider all data sets described before except PRad data which leads to a set of GPDs without any from PRad data. We call this analysis ``Base Fit".
The second analysis is performed by including also the PRad data in the analysis. In this way we can investigate the impact of these data on the final results separately. We call this analysis ``PRad".
In the third analysis, we repeat the second analysis by removing the Mainz data considering the fact that there is a significant tension between them and AMT07 world data~\cite{Ye:2017gyb,Hashamipour:2022noy}. This analysis is called ``PRad\_NoMainz".

Table~\ref{tab:chi2} presents the values of $ \chi^2 $ for the three analyses described above. In the first column, we have listed the data sets with their corresponding observables and references. For each dataset, we have presented the values of $ \chi^2 $ per number of data points, $\chi^2$/$ N_{\textrm{pts.}} $, obtained from the Base Fit, PRad, and PRad\_NoMainz analyses in the second, third, and fourth columns, respectively. In the last row of the table, we have presented the values of total $ \chi^2 $ divided by the number of degrees of freedom, $\chi^2 /\mathrm{d.o.f.} $, for each analysis. As can be seen, the PRad data are in good consistency with other elastic scattering measurements so that including them in the analysis does not change significantly the $ \chi^2 $ of the other data sets. Moreover, removing the Mainz data from the analysis leads to a significant decrease in the value of total $ \chi^2 $ and hence $\chi^2 /\mathrm{d.o.f.} $.
\begin{table}[th!]
\scriptsize
\setlength{\tabcolsep}{7pt} 
\renewcommand{\arraystretch}{1.4} 
\caption{The values of $ \chi^2 $ per number of data points obtained from the Base Fit, PRad, and PRad\_NoMainz analyses. The values of total $ \chi^2 $ divided by the number of degrees of freedom
have been presented in the last row. See text for more details.}\label{tab:chi2}
\begin{tabular}{lccc}
\hline
\hline
  Observable            &  \multicolumn{3}{c}{ $\chi^2$/$ N_{\textrm{pts.}} $  }  \\
                        &     Base Fit    &      PRad     &     PRad\_NoMainz    \\
\hline 		 	
\hline 
Mainz $G_{E}^p$~\cite{A1:2013fsc}                & $275.1 / 77$  & $ 281.6 / 77 $ &    $  -  $ \\
AMT07 $G_E^p/G_D$~\cite{Arrington:2007ux}        & $37.2 / 47$   & $ 37.1 / 47$   &    $ 40.3 / 47 $ \\
AMT07 $R^p $~\cite{Ye:2017gyb}                   & $115.6 / 69$  & $115.7 / 69$   &    $111.7 / 69$ \\
AMT07 $G_{E}^n$~\cite{Ye:2017gyb}                & $30.1 / 38$   & $ 27.1 / 38$   &    $ 25.2 / 38$ \\
AMT07 $G_M^n/\mu_n G_D$~\cite{Ye:2017gyb}        & $32.5 / 33$   & $ 34.4 / 33 $  &    $33.8 / 33 $ \\
GMp12 $\sigma_R$~\cite{Christy:2021snt}          & $16.2 / 13$   & $ 18.4 / 13$   &    $ 13.7 / 13$ \\
PRad $\sigma_R$~\cite{Xiong:2019umf} 1.1 GeV     & $    -    $   & $10.5 / 33$    &    $ 10.2 / 33$ \\
PRad $\sigma_R$~\cite{Xiong:2019umf} 2.2 GeV     & $    -    $   & $45.1 / 38$    &    $ 44.2 / 38$ \\
\hline
Total $\chi^2 /\mathrm{d.o.f.} $                 & $506.7 / 262$ & $569.9 / 333$  &    $279.1 / 256$ \\
\hline
\hline
\end{tabular}
\end{table}

Following the parametrization scan procedure outlined in~\cite{Hashamipour:2021kes,Hashamipour:2022noy,Irani:2023lol}, we identify a set of GPDs characterized by \( \alpha^{\prime}_{g_v^u} = \alpha^{\prime}_{f_v^u} \), \( B_{g_v^u} = 0 \), and \( B_{g_v^d} = 0 \). Notably, releasing these parameters does not result in a reduction of the \( \chi^2 \) value or any improvement in the quality of the fit. The remaining parameters are determined through a standard fit to the data. Table~\ref{tab:par} presents the optimal values of these parameters obtained from the three analyses described above. As evident from the results, the inclusion of the PRad data in the analysis considerably reduces the uncertainties of the parameters, particularly of GPD $ E_v^d $. Additionally, the central values of some parameters exhibit notable changes. On the other hand, removing the Mainz data from the analysis does not substantially alter the final results, particularly for the GPDs \( H_v^q \). However, some variations are observed in the forward limits \( e_v^q(x) \) of the GPDs \( E_v^q \).
\begin{widetext}
\begin{center}
\begin{table}[th!]
\scriptsize
\setlength{\tabcolsep}{8pt} 
\renewcommand{\arraystretch}{1.4} 
\caption{The optimum values of the parameters of the profile functions~(\ref{Eq7}), and distributions $ e_v^q(x) $ of Eq.~(\ref{Eq6}) obtained from three analyses described in the text.}\label{tab:par}
\begin{tabular}{lcccc}
\hline
\hline
 Distribution &  Parameter           &  Base Fit           &  PRad            &  PRad\_NoMainz  \\
\hline 
\hline
$ f_v^u(x) $  & $ \alpha^{\prime} $  & $ 0.745\pm0.016 $ & $ 0.728\pm0.015 $ & $ 0.704\pm0.019 $  \\
			  &	$ A $                & $ 1.017\pm0.053 $ & $ 0.992\pm0.050 $ & $ 0.939\pm0.049 $  \\
			  &	$ B $                & $ 0.772\pm0.049 $ & $ 0.819\pm0.046 $ & $ 0.884\pm0.051 $  \\
\hline
$ f_v^d(x) $  & $ \alpha^{\prime} $  & $ 0.405\pm0.051 $ & $ 0.380\pm0.045 $ & $ 0.350\pm0.053 $  \\
			  &	$ A $                & $ 1.273\pm0.402 $ & $ 1.561\pm0.394 $ & $ 1.559\pm0.490 $  \\
			  &	$ B $                & $ 1.590\pm0.211 $ & $ 1.627\pm0.190 $ & $ 1.683\pm0.226 $  \\
\hline
$ g_v^u(x) $  & $ \alpha^{\prime} $  & $ \alpha^{\prime}_{f_v^u} $ & $ \alpha^{\prime}_{f_v^u} $ & $ \alpha^{\prime}_{f_v^u} $  \\
			  &	$ A $                & $ 0.163\pm0.140 $ & $ 0.318\pm0.153 $ & $ 0.318\pm0.143 $  \\
			  &	$ B $                & $ 0.000         $ & $ 0.000 $         & $ 0.000 $  \\
\hline
$ g_v^d(x) $  & $ \alpha^{\prime} $  & $ 0.606\pm0.258 $ & $ 0.617\pm0.158 $ & $ 0.682\pm0.274 $  \\
			  &	$ A $                & $ 1.429\pm1.514 $ & $ 1.718\pm0.980 $ & $ 1.722\pm1.135 $  \\
			  &	$ B $                & $ 0.000 $         & $ 0.000 $         & $ 0.000 $  \\
\hline
$ e_v^u(x) $  & $ \alpha $           & $ 0.699\pm0.047 $ & $ 0.702\pm0.054 $ & $ 0.679\pm0.033 $  \\
			  &	$ \beta $            & $ 8.178\pm0.484 $ & $ 7.422\pm0.600 $ & $ 7.530\pm0.587 $  \\
			  &	$ \gamma $           & $ 6.555\pm3.107 $ & $ 4.371\pm3.418 $ & $ 2.868\pm1.515 $  \\
\hline
$ e_v^d(x) $  & $ \alpha $           & $ 0.837\pm0.059 $ & $ 0.816\pm0.040 $ & $ 0.813\pm0.062 $  \\
			  &	$ \beta $            & $ 7.189\pm5.065 $ & $ 6.840\pm2.975 $ & $ 6.389\pm3.336 $  \\
			  &	$ \gamma $           & $ 15.626\pm9.326 $ & $ 12.531\pm5.959 $ & $ 15.856\pm11.205 $  \\
\hline 		 	
\hline 	
\end{tabular}
\end{table}
\end{center}
\end{widetext}

After determining the unpolarized valence GPDs $ H_v^q $ and $ E_v^q $ at zero skewness,  now we can calculate the corresponding electromagnetic FFs and hence the charge and magnetic radii of the nucleons using Eq.~(\ref{Eq1}). The results obtained for the proton and neutron charge and magnetic radii from three GPD analyses performed in the present study have been summarized in Table~\ref{tab:radii}. Note that the uncertainties have been calculated by considering also the uncertainties from the \texttt{NNPDF} PDFs (in fact, they have significant contributions). 
\begin{table}[th!]
\scriptsize
\setlength{\tabcolsep}{4pt} 
\renewcommand{\arraystretch}{1.4} 
\caption{The results obtained for the proton and neutron charge and magnetic radii from three GPD analyses performed in the present study.}\label{tab:radii}
\begin{tabular}{lccc}
\hline
\hline

Observable        &     Base Fit    &      PRad     &     PRad\_NoMainz  \\
\hline
\hline
$r_{pE}$             & $ 0.8564 \pm 0.0136 $       & $ 0.8558 \pm 0.0135$       & $ 0.8542 \pm 0.0137 $  \\
$r_{pM}$             & $ 0.8242 \pm 0.0596 $       & $ 0.8268 \pm 0.0533$       & $ 0.8243 \pm 0.0605 $  \\
$<r_{nE}^2>$         & $-0.1132 \pm 0.0272 $       & $-0.1181 \pm 0.0270$       & $-0.1244 \pm 0.0274 $  \\
$r_{nM}$             & $ 0.8330 \pm 0.1136 $       & $ 0.8367 \pm 0.0845$       & $ 0.8426 \pm 0.1230$   \\
\hline
\hline
\end{tabular}
\end{table}

According to the results obtained, either including the PRad data or removing the Mainz data, which both cover the small $ -t $ region, does not lead to a significant change in the values of $ r_{pE} $ and $ r_{pM} $. This clearly shows that the global analysis of GPDs including a wide range of experimental data (covering a wide range of $ -t $) leads to averaged values for the radii of the nucleons comparing with the case where just the data from a certain experiment or a specific observable are analyzed. For example, analyzing the PRad data solely (covering very small $ -t $ region) results in a small value for $ r_{pE} $~\cite{Xiong:2019umf}, while analyzing them besides other data (whether with small or large values of $ -t $) leads to different values (larger) for $ r_{pE} $~\cite{Ridwan:2023ome}. That's exactly why including the PRad data or removing the Mainz data leads to a considerable change in the values of $<r_{nE}^2>$ and $ r_{nM} $. In fact, since in the case of neutron we have fewer data, adding or removing some data points can affect the final results. As can be seen, both $<r_{nE}^2>$ and $ r_{nM} $ have increased (in magnitude) by adding the PRad data to the analysis. They  have even increased more by removing the Mainz data from the analysis. In a quantitative view, e.g., the central value of $<r_{nE}^2>$ has changed about 10 percent from the Base fit analysis to the PRad\_NoMainz analysis or the magnitude of the uncertainty of $ r_{nM} $ has decreased about 34 percent from the Base fit analysis to the PRad analysis. Another point that should be noted is that the large uncertainties of $ r_{nM} $ come from the large uncertainties of GPDs $ E^q_v $ of the neutron which have greater contributions in $ G^n_M $ rather than $ G^n_E $ due to the greater contribution of $ F_2 $ in $ G^n_M $, especially at small values of $ -t $. 
Overall, a significant part of uncertainties comes from the PDF uncertainties. For example, by excluding the PDF uncertainties, the error of $ r_{pE} $ is decreased from 0.0135 to 0.0031 (the reductions are more significant for the charge radii compared to the magnetic radii). 
Note also that the PRad data leads to a significant reduction in the uncertainty of $ r_{nM} $ and its uncertainty becomes again very large after removing the Mainz data of $ G^p_E $. Therefore, one should pay special attention to the fact that how adding or removing the data belonging to the proton can affect the results obtained for the neutron (both in value and uncertainty). So, it seems that analyzing all experimental data related to the radii of the nucleons simultaneously would be preferable.

Figure~\ref{fig:rpE} shows a comparison between the results obtained for the proton charge radius $ r_{pE} $ from three analyses performed here and the corresponding values obtained from various experiments. The right band and the left wider band belong to the CODATA 2014~\cite{Mohr:2015ccw} and CODATA 2018~\cite{Tiesinga:2021myr}, respectively, while the left narrow band has been taken from PDG 2024~\cite{ParticleDataGroup:2024cfk}. For the scattering experiments, we have presented the results of Refs.~\cite{Xiong:2019umf,A1:2010nsl,Zhan:2011ji,Mihovilovic:2019jiz}. The ordinary hydrogen spectroscopy results have been taken from Refs.~\cite{Grinin:2020txk,Bezginov:2019mdi,Fleurbaey:2018fih,Beyer:2017gug,Brandt:2021yor}, while for the muonic hydrogen spectroscopy we have used the results of Refs.~\cite{Pohl:2010zza,Antognini:2013txn}. As can be seen, there is a large gap between some experiments referred to as the proton-radius puzzle. Except for the PRad experiment, other scattering experiments resulted in a large value for $ r_{pE} $ (note that, as discussed before, if PRad data are analyzed besides other scattering data, $ r_{pE} $ would not be as small as the PRad result). Although the older hydrogen spectroscopy results comprise both small and large $ r_{pE} $ regions, two recent measurements~\cite{Grinin:2020txk,Brandt:2021yor} lie inside the gap. In contrast, the muonic hydrogen spectroscopy leads just to small values of $ r_{pE} $. An interesting thing is that our results which have been obtained from the global analyses of GPDs are placed in the middle of the gap. As if the simultaneous analysis of all data leads to an average value for $ r_{pE} $. Anyway, our results are in better consistency with recent hydrogen spectroscopy results. 
\begin{widetext}
\begin{center}
\begin{figure}[!htb]
    \centering
\includegraphics[scale=0.65]{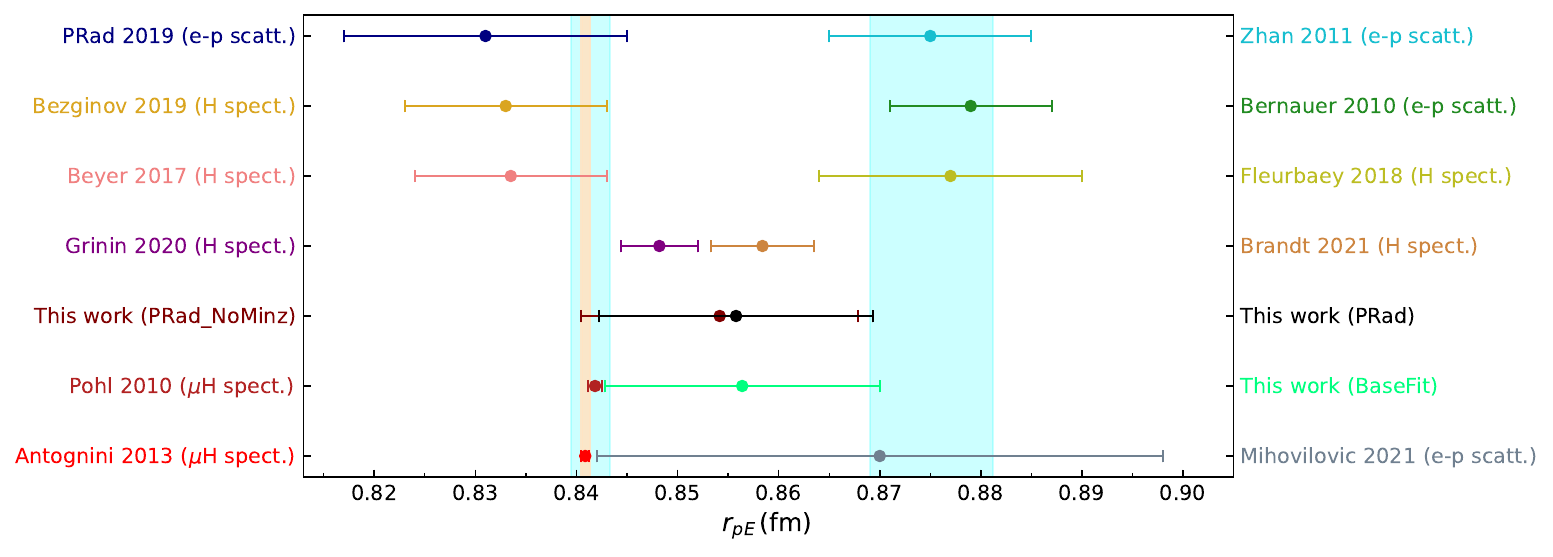}   
    \caption{A comparison between the results obtained for the proton charge radius $ r_{pE} $ from three analyses performed in this study and the corresponding values obtained from various experiments. See text to get more information.}
\label{fig:rpE}
\end{figure}
\end{center}
\end{widetext}

In order to cross check the results obtained and the conclusion reached, we perform a new analysis by including only the PRad data in the analysis and removing other data sets. In this case, the total number of data points included in the analysis would be 71. Such an analysis should basically lead to the same result as PRad experiment for the proton charge radius $ r_{pE} $. Since the PRad data cover only the very small $ -t $ region, there is no need to consider parameters $ A $ and $ B $ in profile functions (they control the medium and large $ -t $ regions). On the other hand, since PRad data can not provide enough constraints on parameters $ \alpha^{\prime} $, we consider them to be equal (note that releasing others does not lead to a significant decrease in the value of $ \chi^2 $ and hence does not improve the quality of the fit). For the forward PDF $ e_v^q $, we use the distributions obtained from the PRad analysis. The value of total $ \chi^2 $ divided by the number of degrees of freedom, $\chi^2 /\mathrm{d.o.f.} $, obtained for this analysis is 0.35 which clearly indicates the good quality of the fit. By calculating the $ r_{pE} $ using the GPDs extracted, one obtains $ r_{pE}= 0.834 \pm 0.008 $ fm. This value is in complete agreement with the PRad experiment $ r_{pE}= 0.831 \pm 0.014 $ fm. This result clearly establishes that considering the medium and large $ -t $  data points as well as other observables in a global analysis of GPDs really changes the extracted value of the $ r_{pE} $. As one includes more data in the analysis from various observables covering wider range of $ -t $, the extracted $ r_{pE} $ becomes more process (or experiment) independent and closer to a unique value.

To be more precise, in our method to determine the nucleon's radii, both large and small $ -t $ data are expected to affect the final results. Please note that the nucleon's radii are also proportional to the profile functions through the derivation of GPDs. This means that even at $ t=0 $ limit, the $ x $-dependence of the profile functions play a crucial role in the values obtained for radii. In other words, the parameters $ A $ and $ B $ in Eq.~(\ref{Eq7}) which are controlled with the medium and large $ -t $ data would be also important. While data belonging to the small values of $ -t $ (such as PRad data) are associated with the parameters $ \alpha^{\prime} $ (note also that the large $ -t $ region is corresponding to the large $ x $ region and vice versa). This is exactly the reason for the importance of both small and large $ -t $ data when one determines the nucleus's radii through the analysis of GPDs.
Actually, since the forward limit $ e_v^q $ and the profile functions $ f_v^q(x) $ and $ g_v^q(x) $ are obtained from whole $ -t $ regions, the medium and large $ -t $ regions are also important and can affect the value of radii. Mathematically, this means that one should first obtain the whole $ -t $ dependency of FFs and then calculate their derivations at $ t=0 $.

\begin{figure}[!htb]
    \centering
\includegraphics[scale=0.7]{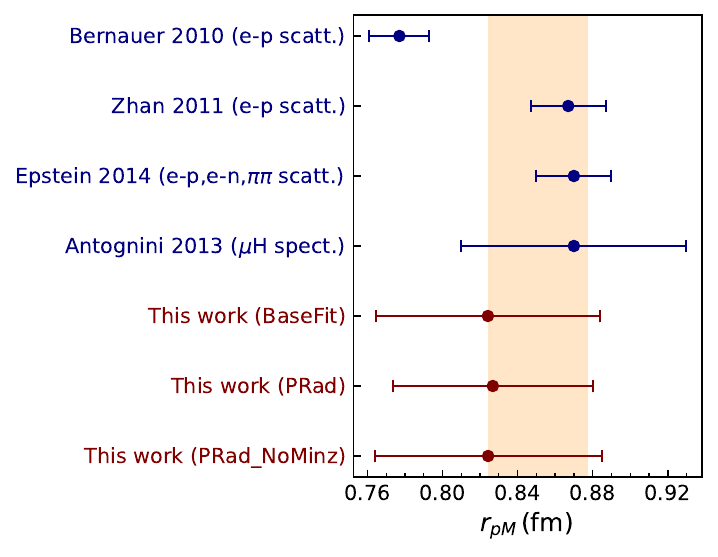}   
    \caption{Same as Fig.~\ref{fig:rpE} but for the proton magnetic radius $ r_{pM} $.}
\label{fig:rpM}
\end{figure}

Figure~\ref{fig:rpM} shows the same comparison as Fig.~\ref{fig:rpE} but for the proton magnetic radius $ r_{pM} $. In this case, there is less information from the experiment compared with the measurements of $ r_{pE} $. We have presented three results from the scattering experiments~\cite{A1:2010nsl,Zhan:2011ji,Epstein:2014zua} and only one result from the muonic hydrogen spectroscopy~\cite{Antognini:2013txn}. Note that the value presented by Epstein \textit{et al.} has been extracted in a model independent way from the electron-proton scattering data in addition to the electron-neutron scattering and $ \pi\pi \rightarrow N\bar{N} $ data. The band shows the corresponding value taken from PDG 2024~\cite{ParticleDataGroup:2024cfk}. As can be seen, in this case, the results of different experiments are in better consistency with each other compared with $ r_{pE} $, especially considering the uncertainties. For example, in Fig.~\ref{fig:rpE} there is a large gap between the results of Zhan 2011~\cite{Zhan:2011ji} and Antognini 2013~\cite{Antognini:2013txn} for $ r_{pE} $ referring to the proton radius puzzle, while the corresponding values for $ r_{pM} $ from these two experiments are in complete agreement. An interesting thing is that our results are almost the average of other data like the case of $ r_{pE} $. 

\begin{figure}[!htb]
    \centering
\includegraphics[scale=0.68]{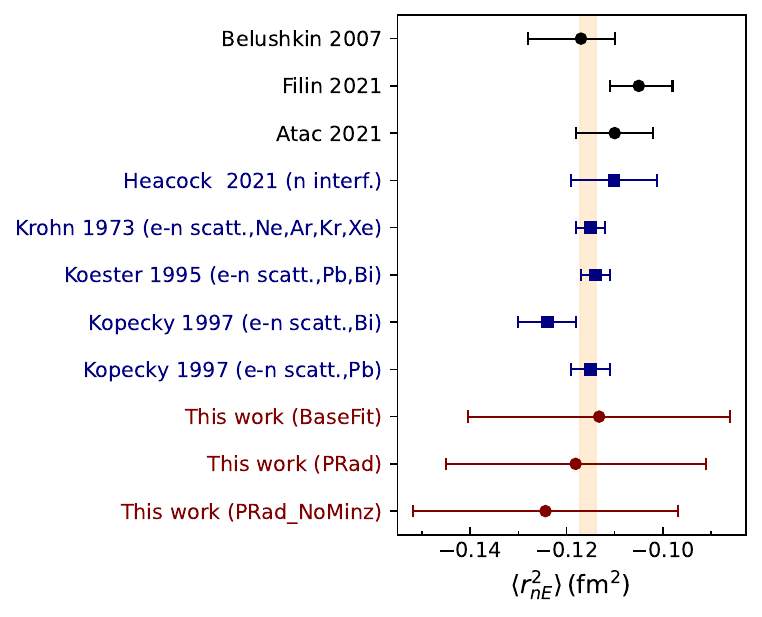}   
    \caption{A comparison between the results obtained for $<r_{nE}^2>$ from three analyses performed in this study and the corresponding values obtained from various experiments analyses. See text to get more information.}
\label{fig:rnE}
\end{figure}

The results obtained for $<r_{nE}^2>$ have been presented in Fig.~\ref{fig:rnE} and compared with the available experimental measurements~\cite{Heacock:2021btd,Krohn:1973re,Koester:1995nx,Kopecky:1997rw} (square symbols) and the corresponding values obtained from various analyses including the dispersion analysis of the nucleon electromagnetic FFs data~\cite{Belushkin:2006qa}, chiral effective field theory~\cite{Filin:2020tcs}, and the extraction of the neutron electric FF from the nucleon-to-delta ($ N\rightarrow \Delta $) quadrupole transitions~\cite{Atac:2021wqj}. As before, the band shows the value taken from PDG 2024. The experimental measurements mostly belong to the electron-neutron scattering experiments through the scattering off different nuclei. However, there is a measurement~\cite{Heacock:2021btd} from an unusual method called Pendell\"osung interferometry where the momentum dependence of the structure factors enabled the researchers to measure the charge radius of the neutron. As can be seen, all results are in good consistency with each other. As mentioned before, the large uncertainties in our results come from our poor knowledge of the neutron GPDs due to the lack of data in this case (note that the uncertainties include also the PDFs uncertainties which are relatively large, especially in the case of \texttt{NNPDF}). Actually, it is vital to measure the neutron electromagnetic FFs precisely to put strong constraints on the forward limit of the down quark GPD $ E_v^d $ and its profile function to reduce the uncertainty of the charge and magnetic radii of the neutron. Note that the radii are related also to the profile functions through the derivation of the exponential in ansatz~(\ref{Eq5}).

\begin{figure}[!htb]
    \centering
\includegraphics[scale=0.7]{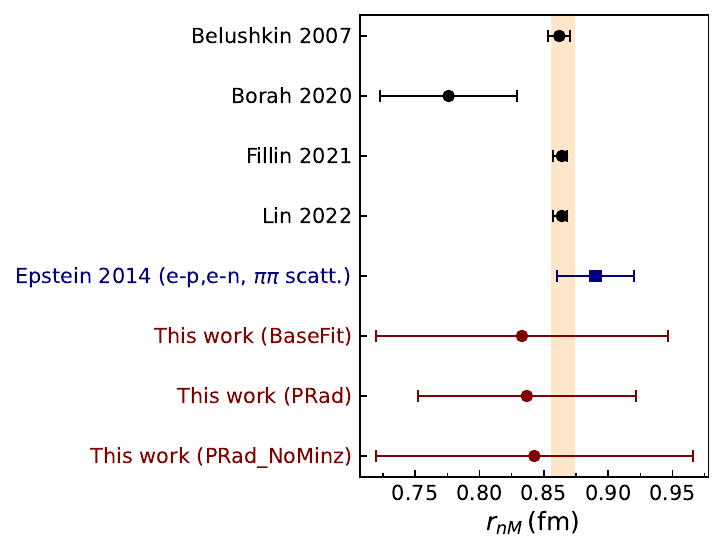}   
    \caption{Same as Fig.~\ref{fig:rnE} but for the neutron magnetic radius $ r_{nM} $.}
\label{fig:rnM}
\end{figure}

Figure~\ref{fig:rnM} shows the same comparison as Fig.~\ref{fig:rnE} but for the neutron magnetic radius $ r_{nM} $. In this case, there is only one measurement by Epstein \textit{et al.} from the scattering experiments. In addition to the results of Belushkin 2007~\cite{Belushkin:2006qa} and Filin 2021~\cite{Filin:2020tcs} described before, here we have also presented the results of two other analyses, namely, Borah 2020~\cite{Borah:2020gte} and Lin 2022~\cite{Lin:2021xrc}. The former has been obtained
from a global fit to electron scattering data and precise charge radius measurements, while the latter has been obtained from a combined analysis of the electromagnetic FFs of the nucleon in the space- and timelike regions using dispersion theory. According to this figure, except for Borah 2020, all results are in very good agreement with each other and also the corresponding value taken from PDG 2024 (the band). Note that among the results obtained from the present study, the analysis called PRad that contains all experimental data sets (including the PRad and Mainz data) has a smaller uncertainty. This clearly indicates that inducing more precise data, especially the neutron data, in the analysis of GPDs can significantly reduce the large uncertainties observed in Figs.~\ref{fig:rnE} and~\ref{fig:rnM}. Therefore,  ongoing and upcoming experiments such as the MUSE experiment at PSI, the COMPASS++/AMBER experiment at CERN, the PRad-II experiment at JLab, and future electron scattering experiments at Mainz, and the UL$ Q^2 $ experiment at Tohoku University can play a crucial role in this regard. See Refs.~\cite{Karr:2020wgh,Gao:2021sml} to get more information. The proton charge radius can even be measured through dimuon photoproduction off a proton target~\cite{Lin:2024bzo}.
%

	
	\textit{\textbf{\textcolor{violet}{Conclusions}}}~~
In this study, we determined the charge and magnetic radii of the proton and neutron through a global analysis of GPDs at zero skewness for the first time. To this aim, we first extracted the unpolarized valence GPDs $ H_v^q $ and $ E_v^q $ with their uncertainties from a $ \chi^2 $ analysis of a wide range of the experimental data of the nucleons electromagnetic FFs and the elastic electron-proton reduced cross section. Then, considering the fact that the nucleon's radii are the slope of the electromagnetic FFs at $ Q^2=0 $, we calculated the proton and neutron charge and magnetic radii. The results related to the analysis that contains all experimental data sets, namely PRad, are as follows: 
\begin{align}
r_{pE} =& 0.8558 \pm 0.0135~\textrm{fm}\,, \nonumber \\ 
r_{pM} =& 0.8268 \pm 0.0533~\textrm{fm}\,, \nonumber \\ 
\left<r_{nE}^2\right> =& -0.1181 \pm 0.0270~\textrm{fm}^2\,, \nonumber \\    
r_{nM} =& 0.8367 \pm 0.0845~\textrm{fm}\,.
\label{Eq17}
\end{align}
It should be noted that the uncertainties include also the PDFs uncertainties which are relatively large. Overall, the neutron radii have larger uncertainties than the corresponding values for the proton due to the lack of information about the neutron. In this way, the ongoing and upcoming experiments~\cite{Karr:2020wgh,Gao:2021sml} will play definitely a crucial role to shed light on this issue. For instance, it has been demonstrated now that the two-photon exclusive production of lepton pairs at the Electron-Ion Collider can provide a unique opportunity to measure the proton's elastic electromagnetic FFs and, consequently, the nucleon's radii~\cite{Chwastowski:2022fzk}.

In the context of the proton-radius puzzle, our results are placed in the middle of the gap between various experimental measurements obtained from the scattering experiments and hydrogen spectroscopy. This means that the simultaneous analysis of all data leads to an average value for $ r_{pE} $ and also $ r_{pM} $ extracted so far. In the case of $ r_{pE} $, our results are in better consistency with recent ordinary hydrogen spectroscopy result~\cite{Grinin:2020txk,Brandt:2021yor}. For the neutron, the radii obtained from our analyses are in good consistency with the results of other experiments and analyses as well as the PDG 2024~\cite{ParticleDataGroup:2024cfk}. 

As an important result, we showed that the global analysis of GPDs considering a wide range of the experimental data (covering a wide range of $ Q^2 $) ensures robust and consistent results for the radii of the nucleons, avoiding values that are either too small or too large, while analyzing the data from just a certain experiment or a specific observable leads to different results. We indicated that including or removing data belonging to the proton can affect the results obtained for the neutron (both in value and uncertainty). We emphasize the importance of a simultaneous analysis of all available experimental data related to nucleon radii, rather than relying on individual experiments, specific observables, or limited kinematic regions. Actually, this method has an advantage that it can take into account the possible correlations between the $ G_E $ and $ G_M $ data of the proton as well as those between the proton and neutron data.


	\textit{\textbf{\textcolor{violet}{Acknowledgements}}}~~
We thanks Weizhi Xiong and Ashot Gasparian for providing us with the PRad data. 
M. Goharipour is thankful to the School of Particles and Accelerators, Institute for Research
in Fundamental Sciences (IPM), for financial support provided for this research. F. Irani and K. Azizi are thankful to Iran National Science Foundation (INSF) for financial support provided for this research under grant No. 4033039.
	
	\onecolumngrid

	\twocolumngrid


	\onecolumngrid

\end{document}